\documentclass[pre]{revtex4}
\usepackage{setspace}
\usepackage{graphicx}
\usepackage{dcolumn}
\usepackage{amsmath}
\usepackage{amsfonts}
\usepackage{amssymb}
\usepackage{subfigure}
\usepackage{color}
\begin{document}

\title{Emergence of Dynamic Cooperativity in the Stochastic Kinetics of Fluctuating Enzymes}

\author{Ashutosh Kumar}
\author{Sambarta Chatterjee}
\altaffiliation{Department of Chemistry, University of Illinois, 600 S. Goodwin Avenue, Urbana, Illinois 61801, USA}
\author{Mintu Nandi}
\altaffiliation{Department of Chemistry, University of Calcutta, 92 APC Road, Kolkata-700009, India}
\author{Arti Dua}
\email{arti@iitm.ac.in}

\affiliation{Department of Chemistry, Indian Institute of Technology, Madras, Chennai-600036, India}



\date{\today}

\begin{abstract}
\noindent
Dynamic cooperativity in monomeric enzymes is characterized in terms of a non-Michaelis-Menten kinetic behaviour. The latter is believed to be associated with mechanisms that include multiple reaction pathways due to enzymatic conformational fluctuations. Recent advances in single-molecule fluorescence spectroscopy have provided new fundamental insights on the possible mechanisms underlying reactions catalyzed by fluctuating enzymes.  Here, we present a bottom-up approach to understand enzyme turnover kinetics at physiologically relevant mesoscopic concentrations informed by mechanisms extracted from single-molecule stochastic trajectories. The stochastic approach, presented here, shows the emergence of dynamic cooperativity in terms of a slowing down of the Michaelis-Menten (MM) kinetics  resulting in negative cooperativity. For fewer enzymes, dynamic cooperativity emerges due to the combined effects of enzymatic conformational fluctuations and molecular discreteness. The increase in the number of enzymes, however, suppresses the effect of enzymatic conformational fluctuations such that dynamic cooperativity emerges solely due to the discrete changes in the number of reacting species. These results confirm that the turnover kinetics of fluctuating enzyme based on the parallel-pathway MM mechanism  switches over to the single-pathway MM mechanism with the increase in the number of enzymes. For large enzyme numbers, convergence to the exact MM equation occurs in the limit of very high substrate concentration as the stochastic kinetics approaches the deterministic behaviour. 
\end{abstract}
\maketitle

\pagebreak

\section{Introduction}

Several monomeric enzymes show deviation from the classical Michaelis-Menten (MM) expression that quantifies the initial rate of product formation in enzyme catalyzed reactions.$^{1-5}$ The MM equation is expressed in terms of the steady-state enzyme velocity, $v = \frac{k_2 [S][E]_0}{K_M + [S]}$, where $[S]$ is the substrate concentration, $[E]_0 = [E] + [ES]$ is the initial enzyme concentration and $K_M = (k_2+k_{-1})/k_1$ is the Michaelis constant that depends on the rate constants for substrate binding ($k_1$), substrate release ($k_{-1}$)  and product formation ($k_2$) in the MM mechanism:$^{6,7}$
\begin{equation}\label{mmm}
 E + S \mathop{\rightleftharpoons}^{k_1}_{k_{-1}} ES  \xrightarrow{k_2}  E + P, 
\end{equation}
The enzyme velocity in the MM equation represents a non-cooperative kinetic response of  monomeric enzymes.  Any deviation from the latter is characterized in terms of a non-Michaelis-Menten kinetic behaviour, which is a manifestation of dynamic (or kinetic) cooperativity of monomeric enzymes. These deviations correspond to positive or negative cooperativity that result in either a speeding up or slowing down of the MM kinetics with the increase in the substrate concentration.$^{1-5}$

Although enzyme cooperativity is traditionally linked to enzymes with multiple binding sites, where binding of a substrate at a site can influence the time scale of substrate binding at another site,$^{1}$  the emergence of dynamic cooperativity in several monomeric enzymes is believed to be associated with mechanisms that include multiple reaction pathways.$^{2,8-11}$ These multiple reaction pathways are  due to the presence of enzyme conformational fluctuations in the reaction mechanism. In several earlier kinetic studies,$^{9-11}$  the parallel-pathway MM mechanism that includes the effects of enzymatic conformational fluctuations in terms of interconversions between two enzyme and enzyme-substrate conformers has been proposed as a minimal model to observe the effects of dynamic cooperativity.  The latter is given by
\begin{eqnarray}\label{ppm}
& &E_1 + S  \mathop{\rightleftharpoons}^{k_{11}}_{k_{-11}} ES_1 \xrightarrow{k_{21}}   E_1^0 + P, E_1^0 \xrightarrow{\delta_{21}} E_1 \nonumber\\
& & {\alpha} {\upharpoonleft \downharpoonright}{\alpha}~~~~~~~~~ {\beta} {\upharpoonleft \downharpoonright}{\beta}~~~~~~ {\gamma} {\upharpoonleft \downharpoonright}{\gamma}\nonumber\\
& & E_2 + S \mathop{\rightleftharpoons}^{k_{12}}_{k_{-12}}  ES_2 \xrightarrow{k_{22}}  E_2^0 +P,  E_2^0 \xrightarrow{\delta_{22}} E_2 
\end{eqnarray} 
In the above reaction mechanism, each enzyme, enzyme-substrate  and regenerated enzyme state can interconvert between two conformational states specified by $E_{1} \rightleftharpoons E_{2}$, $ES_{1} \rightleftharpoons ES_{2}$ and $E_{1}^0 \rightleftharpoons E_{2}^0$ with the rate constants $\alpha$, $\beta$ and $\gamma$ respectively. The reaction step $E_{i}^0 \rightarrow E_{i}$ is assumed to occur instantaneously so that $E_i$ and $E_i^0$ states are effectively identical.$^{12-17}$ Interestingly, several recent studies$^{12-14,18-20}$ have proposed the parallel-pathway mechanism [Eq. (\ref{ppm})] as a minimal model to rationalize the results of single-molecule kinetic measurements$^{12,13}$ that observe deviations from the MM kinetics due to significant temporal fluctuations in the reaction pathway.  The latter, termed as dynamic disorder, is a result of multiple competing steps in the parallel-pathway mechanism.

In the context of enzyme kinetics, the term ``cooperativity'' implies  a collective response  either in terms of multiple binding sites in a single enzyme or in terms of temporal fluctuations in multiple copies of a monomeric enzyme due to interconversions between different conformer states. In enzyme catalyzed reactions, however, the enzyme concentration is not thermodynamically large. Typical enzyme concentration  varies between picomolar to nanomolar in {\it vitro} reactions and nanomolar to micromolar in {\it vivo} reactions.$^{21,22}$ The substrate concentration, in contrast, is  typically million times higher in the former case and ten to ten thousand times higher in the latter case. Such low concentration, lying between the extremes of thermodynamic large and single-enzyme limits, correspond to mesoscopic concentration regimes. At mesoscopic concentration, enzymes are discrete entities, which as a result of random reversible binding and unbinding with substrates  undergo discrete integer increment or decrement  in their numbers. While the randomness stems from the  incessant thermal fluctuations in a reacting system, it manifests as the random and discrete integer jumps in the time evolution of the number of reacting species. The latter is a manifestation of intrinsic stochasticity due to molecular discreteness at mesoscopic concentrations.$^{23-29}$ This, combined with the inherently probabilistic nature of interconversions between different enzymatic conformers in the parallel-pathway mechanism,  necessitates the use of stochastic kinetic approaches to understand the emergence of dynamic cooperativity in monomeric enzymes.  Earlier studies have addressed this problem solely in terms of  deterministic kinetics.$^{9-11}$

In this work, we present a stochastic approach to understand the enzyme turnover kinetics at physiologically relevant mesoscopic concentrations informed by mechanisms extracted from single-molecule stochastic trajectories.$^{12-14,18-20}$ The bottom-up approach, based on the superposition of renewal processes,  exploits the renewal nature of the waiting time distribution of a single enzyme to obtain the waiting time distribution of $N$ (independent and identical) enzyme copies as a single pooled output.$^{30}$ The latter is verified against exact stochastic simulations based on the Doob-Gillespie algorithm.$^{31-34}$ {The calculation of the waiting time distribution for the first product formation  belongs to a general  class of  first-passage problems in stochastic processes.$^{35-37}$  A recent theoretical work based on a first-passage analysis, for instance, has shown that waiting time distribution of a single molecule undergoing transition from one arbitrary state to another in a complex network contain complete information on number of intermediate states, pathways, and transitions between initial and final states.$^{38}$ The advantage of the latter is that structural and dynamical information of the underlying complex networks can be predicted by analysing distributions of events without any assumption of a specific model. A recent review article has shown that the dimensionless variance (randomness parameter) of the waiting time distribution for the first product formation in single-enzyme kinetics can provide structural and dynamical information hidden in several different enzymatic networks of varying complexity.$^{18}$  While the present work considers a specific mechanism [Eq. (\ref{ppm})] to calculate the waiting time distribution for $N$ enzyme copies starting from the waiting time distribution of a single enzyme, it presents a general method that can be used to calculate the waiting time distributions between two arbitrary events in a complex network of $N$ molecules from the waiting time distribution of a single molecule. This can, then, provide an opportunity  to understand the emergent collective behaviour within a more general framework based on first-passage approaches.}

\section{Modelling turnover kinetics of fluctuating enzymes at mesoscopic concentrations}

Kinetics of thermodynamically large number of enzymes and substrates is well described by classical deterministic kinetics that provide temporal variation of concentration of reactants, intermediates and products in bulk, with concentration being the unit of continuous transformation.$^{7}$  However, at mesoscopic concentrations the enzyme turnover kinetics as described by smooth and continuous mean product formation rate is not valid.  Instead, the number of products in each enzymatic turnover cycle increases in discrete integer jumps, occurring randomly in time, such that the the waiting time between two consecutive turnover events are stochastic quantities.$^{35,36}$ The turnover kinetics at mesoscopic concentration is, thus, described by the waiting time distribution and its moments - the mean waiting time and the randomness parameter. { Recent single-molecule kinetic measurements and theoretical approaches have quantified the effects of conformational fluctuations in the enzyme turnover kinetics in terms of the waiting time distributions and its first two moments.$^{12-15,39-44}$ }


Assuming that the system is well-mixed, the stochastic turnover kinetics at mesoscale is exactly described by the chemical master equation (CME) approach of stochastic processes.$^{35,36,45,46}$ The CME describes the time evolution of the joint probability distribution  $P(n_{E_1}, n_{ES_1}, n_{E_2}, n_{ES_2}, n_{E_1^0}, n_{E_2^0}, n_P; t)$ in terms of the CME transition probabilities for each step in the parallel-pathway MM mechanism [Eq. {\ref{ppm}]. The CME accounts for the intrinsic stochasticity of an enzymatic reaction due to the discrete nature of the reacting species and fluctuations in their conformational states. The resulting equation is
\begin{eqnarray}\label{cme-ppm}
 {\partial_{t} P} &=& 
   \left[ \sum_{i=1}^2 \left( k_{1i}' (\mathbb{E}_{E_i} \mathbb{E}_{E_iS}^{-1} - 1 \right) n_{E_i} +  k_{-1i} (\mathbb{E}_{E_i}^{-1} \mathbb{E}_{E_iS} - 1) n_{E_iS} + k_{2i} (\mathbb{E}_{E_iS} \mathbb{E}_{E_p}^{-1} \mathbb{E}_{E_i^0}^{-1} - 1) n_{E_iS} \right.  \nonumber\\
& & + \left. \left.  \delta_{2i} (\mathbb{E}_{E_i^0} \mathbb{E}_{E_i}^{-1}  - 1) n_{E_i^0} \right)   \right] P +  \alpha \left[ (\mathbb{E}_{E_1} \mathbb{E}_{E_2}^{-1} - 1) n_{E_1} + (\mathbb{E}_{E_1}^{-1} \mathbb{E}_{E_2} - 1) n_{E_2} \right] P \nonumber\\
& & +  \beta \left[ (\mathbb{E}_{E_1S} \mathbb{E}_{E_2S}^{-1} - 1) n_{E_1S} + (\mathbb{E}_{E_1S}^{-1} \mathbb{E}_{E_2S} - 1) n_{E_2S} \right] P \nonumber\\
& & +  \gamma \left[ (\mathbb{E}_{E_1^0} \mathbb{E}_{E_2^0}^{-1} - 1) n_{E_1^0} + (\mathbb{E}_{E_1^0}^{-1} \mathbb{E}_{E_2^0} - 1) n_{E_2^0} \right] P.
 \end{eqnarray}
where $k_{1i}' = k_{1i}[S]$ is the pseudo first order rate constant, $\mathbb{E}^{\pm}$ is a step operator which yields\\ $\mathbb{E}^{\pm} \left[{n_j} P(n_j; t)\right] = (n_j \pm 1) P(n_j \pm 1; t)$.$^{35}$

In the absence of enzyme conformational fluctuations, the CME for the Michaelis-Menten [Eq. {\ref{mmm}] mechanism can be directly written as
\begin{eqnarray}\label{cme-mmm}
\partial_{t} P(n_E, n_{ES}, n_P; t) &=& \left[ k_{1}' (\mathbb{E}_{E} \mathbb{E}_{ES}^{-1} - 1 ) n_{E} +  k_{-1} (\mathbb{E}_{E}^{-1} \mathbb{E}_{ES} - 1) n_{ES} \right.  \nonumber\\
& & \left. +  k_{2} (\mathbb{E}_{ES} \mathbb{E}_{E}^{-1} \mathbb{E}_{P}^{-1}  - 1) n_{ES}   \right] P(n_E, n_{ES}, n_P; t),
\end{eqnarray}
where $k_1' = k_1 [S]$. The above equation accounts for stochasticity due to molecular discreteness in the MM kinetics in terms of the discrete change in the number of reactants, intermediates and products. 

In a previous work,$^{25}$, the CME [Eq. (\ref{cme-mmm})] for the MM mechanism [Eq. (\ref{mmm})] for $N$ enzyme copies was solved exactly using the generating function method, where  $N = n_{E} + n_{ES}$ at all times. The multivariate nature of the joint probability distribution required a formal connection between the counting and point process description of stochastic processes.$^{47,48}$ Using the latter, an exact expression for the first waiting (or turnover) time distribution was obtained for $N$ (independent and identical) enzyme copies, the final form of which is given by
\begin{equation}\label{wtd1}
w(\tau_1; N) = \frac{k_2 k_1' N}{(2 A)^N} \left[e^{(A-B)\tau_1} - e^{-(A+B)\tau_1}\right] \left[(A+B) e^{(A-B)\tau_1} + (A-B) e^{-(A+B) \tau_1}\right]^{N-1},
\end{equation}
where $A = \frac{1}{2} \sqrt{(k_1'+k_{-1}+k_2)^2 - 4 k_1' k_2}$ and $B = \frac{1}{2} (k_1' + k_{-1} + k_2)$.


 It is to be noted that Eq. (\ref{cme-ppm}) is a linear master equation. In a recent work, an exact solution for a general class of linear master equations was obtained in terms of multi-Possionian distributions with time-dependent expectation values.$^{49}$ It is possible, in principle, to use this exact result, together with the formula for temporal waiting time distributions derived in $^{25}$ to obtain the quantities required in the present work. However, the marginalization over the multiple intermediate variables, a necessary step in the use of formula for the temporal waiting time distribution,  requires an involved combinatoric calculation. In what follows, we circumvent this by employing an approach based on the superposition of renewal processes (henceforth called SRP) to calculate the first waiting time distribution.$^{30}$ This method  exploits the renewal nature of the turnover statistics of a single enzyme to yield the exact waiting time distribution for independent and identical $N$ enzyme copies.

%

We begin by establishing the renewal nature of the turnover kinetics of a single enzyme (in the presence of conformational fluctuations) using exact stochastic simulations. We then use the exact expression for the first waiting time  distribution of a single enzyme in the absence of conformational fluctuations, Eq. (\ref{wtd1}), to show that the first waiting time distribution obtained from the method of SRP is in exact agreement with the generating function method.  Having established the latter, the method of SRP is used to obtain an analytical expression for the waiting time distribution of $N$ fluctuating enzymes. The latter expression is verified against exact stochastic simulations.

\subsection{Renewability of turnover kinetics at the single-enzyme level}

Stochastic trajectories obtained from a CME typically yield the time evolution of the number of reacting species in a given reaction mechanism. The stochastic trajectories can be analysed using the count and point process description of stochastic processes.$^{47,48}$ While the counting process description counts the number of products formed in a given time, the point process description  yields the time for the $p$th enzymatic turnover $T_p$ such that $T_0 = 0$.  The difference between the consecutive turnover times yield the $p$th waiting time, $\tau_p = T_p - T_{p-1}$ for $p = 1, 2, 3, \cdots$. The latter implies that the first turnover time distribution, $f(T_1;N)$, is the same as the first waiting time distribution $w(\tau_1; N) = f(T_1; N)$. 

In order to classify the enzyme turnover statistics as renewal in nature, we generate exact numerical trajectories of the CME [Eq. (\ref{cme-ppm})] using the Doob-Gillespie algorithm $^{31-33}$. Stochastic simulations are performed using StochPy libraries.$^{34}$ Each stochastic trajectory is obtained for a sufficiently long time interval such that many turnover events are included. We generate ensembles of typically $10^6$ trajectories to obtain the $pth$ turnover and waiting times. The latter is used to obtain the waiting time distribution, $w(\tau_p, N)$, for the $p$th turnover. Figs (1) shows the $p$th waiting time distribution for $p = 1, 10, 100$ turnover events for $N=1$ enzyme. For a single enzyme, the waiting time distributions for different  turnover numbers collapse onto each other. The latter shows $w(\tau_1; N=1) = w(\tau_p; N=1)$ for $p = 2, 3, \cdots$, implying that the waiting times $\tau_p$ are identically distributed. This establishes the renewal nature of the turnover statistics of a single fluctuating enzyme. Interestingly, the multiexponential decay of the waiting time distribution in the presence of enzyme conformational fluctuations does not affect the renewal nature of the turnover statistics as the waiting times remain identically distributed. { The details of stochastic simulations, the choice of parameters and renewal nature of the single-enzyme turnover kinetics is discussed further in Supplementary Information, SI(I).}


Having established the renewal nature of the turnover kinetics of a single fluctuating enzyme, we use the waiting time distribution for a single enzyme to obtain the waiting time distribution at mesoscopic number of enzymes. The details are presented in the next section.

\subsection{A bottom-up approach based on the superposition of renewal processes}

\begin{figure}[t]
\centering
\vspace{-1.00in}
\includegraphics[scale=.8]{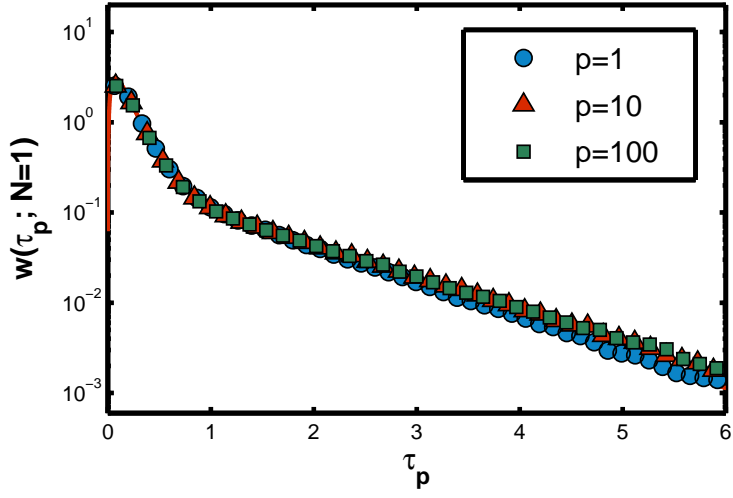}
\vspace{-1.5in}
\caption{Temporal variation of the waiting time distribution (in non-dimensional units) for the parallel-pathway MM mechanism obtained from stochastic simulations.  For a single enzyme, the waiting time distributions for $p = 1, 10,100$ enzyme turnovers are identically distributed implying that the turnover kinetics is renewal in nature.   Parameter values are $k_{11} = k_{-11} = k_{12} = k_{-12} = 0.5$, $k_{21} = 5$, $k_{22} = 0.1$, $\alpha = \beta = \gamma = 1$, $\delta_{21} = \delta_{22} = 1000$ and $[S] = 50$.}
\end{figure}

In this section, we present a bottom-up approach to obtain the waiting time distribution for $N$ enzyme copies.  This approach, based on the superposition of renewal  processes (SRP), was first introduced by Cox and Smith to understand the statistical properties of discrete neuron signals  in neurophysiology.$^{30}$ The SRP method exploits the renewal nature of statistical properties of the sequence of impulses from an individual neuron to obtain a single pooled output. The latter corresponds to the combined sequence of pulses at the central cell. 

In the previous section, the renewal nature of the waiting time distribution of a single enzyme was established using $w(\tau_1; N=1) = w(\tau_p; N=1)$, which showed that the waiting times, $\tau_p$, are identically distributed. Below, we use the waiting time expression for a single enzyme, $w(\tau_1, N=1)$ to obtain a single pooled output corresponding to the waiting time distribution of $N$ enzyme copies, $w(\tau_1; N)$. The resulting expression, as given by the SRP method of Cox and Smith,$^{30}$ is given by 
\begin{equation}\label{srp}
w(\tau_1; N) = N ~ w(\tau_1; N=1) \left(\int_{\tau_1}^\infty~ w(\tau_1'; N = 1)~d\tau_1'\right)^{N-1}
\end{equation}

It is to be noted that the exact analytical expression for the waiting time distribution for $N$ enzyme copies in the absence of conformational fluctuations [Eq. (\ref{wtd1})] is known from the solution of the CME [Eq. (\ref{cme-mmm})] using the generating function method. We first obtain  Eq. (\ref{wtd1}) using the method of SRP [Eq. (\ref{srp})]. This requires the analytical expression for the waiting time distribution of a single enzyme in the absence of conformational fluctuations. The latter   can be easily obtained as the CME [Eq. (\ref{cme-mmm})] for a single enzyme reduces to a set of coupled differential equations. {While the details are given in Ref. (14), the final expression is}
\begin{equation}\label{wtd2}
w(\tau_1; N=1) = \frac{k_2 k_1[S]}{(2 A)} \left[e^{(A-B)\tau_1} - e^{-(A+B)\tau_1}\right].
\end{equation}
Eq. (\ref{wtd2}) when substituted into Eq. (\ref{srp}) yields the waiting time distribution for $N$ enzyme copies, $w(\tau_1; N)$. The resulting expression is the same  as Eq. (\ref{wtd1}), obtained from the generating function method.

A variant of the SRP method$^{50}$ has been used earlier to pool the {\it steady-state} waiting time distribution of a single enzyme. It, however, yields an expression which is different from the {\it exact} analytical expression obtained from the generating function method [Eq. (\ref{wtd1})].  The latter, as is shown here, can be {\it {exactly}} recovered from the general method of SRP, Eq. (\ref{srp}), given by Cox and Smith.$^{30}$ 

The method of superposition of renewal processes, thus, provides a simple approach to obtain an exact analytical expression for the waiting time distribution at mesoscopic concentration. This is particularly useful in cases where an analytical solution of the CME  is difficult to obtain. In the next section, we use the method of SRP to obtain the waiting time distribution for $N$ independent and identical copies of a fluctuating enzyme. The latter implies that each enzyme can interconvert between two conformational states based on the parallel-pathway mechanism.

\subsection{Waiting time distribution for turnover kinetics of  fluctuating enzymes at mesoscopic concentrations}

Having established the renewal nature of the turnover kinetics of a single fluctuating enzyme [Fig (1)], we can use the expression for the waiting time distribution of a  single fluctuating enzyme $w(\tau_1; N=1)$ in Eq. (\ref{srp}) to obtain the waiting time distribution of $N$ fluctuating enzyme copies, $w(\tau_1; N)$. {For a single enzyme, the waiting time distribution for the first product formation in the presence of conformational fluctuations, $w(\tau_1; N=1)$, can be obtained from the CME . This is done by deducing the differential equations that describe the time evolution of $P_{E_1}$, $P_{ES_1}$, $P_{E_1^0}$, $P_{E_2}$,$P_{ES_2}$ and $P_{E_2^0}$ from the CME [Eq. (\ref{cme-ppm})]. While the details are given in Supplementary Information, SI(II), the resulting expression  is}
\begin{eqnarray}\label{wtd1-ppm}
\lefteqn{w(\tau_1; N=1) =}\nonumber\\
& & \frac{1}{(a-b)(c-d)} \left[ \frac{(c-d)~ \eta_a ~e^{-a\tau_1}}{(a-c)(a-d)}  - \frac{(c-d)~\eta_b ~e^{-b\tau_1}}{(b-c)(b-d)} + \frac{(a-b) ~\eta_c ~e^{-c\tau_1}}{(c-a)(c-b)}  - \frac{(a-b)~ \eta_d ~e^{-d \tau_1}}{(d-a)(d-b)}\right],\nonumber\\
\end{eqnarray}
where $\eta_a = {a\mu_A - u_B - a^2 k_{21} k_{11} [S]}$, $\eta_b = {b\mu_A - u_B - b^2 k_{21} k_{11} [S]}$,  $\eta_c = {c\mu_A - u_B - c^2 k_{21} k_{11} [S]}$ and  $\eta_d = {d\mu_A - u_B - d^2 k_{21} k_{11} [S]}$ with $\mu_A = k_{21} A_1 + k_{22} A_2$ and $\mu_B =  k_{21} B_1 + k_{22} B_2$. In the above equations, $a$, $b$, $c$ and $d$ are the effective rate constants which are the solutions of the quartic equation $s^4 + \lambda_1 s^3 + \lambda_2 s^2 + \lambda_3 s + \lambda_4 = 0$ and $A_1$, $A_2$, $B_1$, $B_2$, $\lambda_1$, $\lambda_2$, $\lambda_3$ and $\lambda_4$, have complicated dependence on the rate constants, whose explicit expressions are given in Supplementary Information, SI(II).

To obtain the waiting time distribution for $N$ fluctuating enzyme copies, Eq. (\ref{wtd1-ppm}) is substituted into Eq. (\ref{srp}). After simplification, the final expression is given by
\begin{eqnarray}\label{wtdn-ppm}
\lefteqn{w(\tau_1; N) =}\nonumber\\
& & \frac{N}{((a-b)(c-d))^N} \left[ \frac{(c-d)~ \eta_a ~e^{-a\tau_1}}{(a-c)(a-d)}  - \frac{(c-d)~\eta_b ~e^{-b\tau_1}}{(b-c)(b-d)} + \frac{(a-b) ~\eta_c ~e^{-c\tau_1}}{(c-a)(c-b)}  - \frac{(a-b)~ \eta_d ~e^{-d \tau_1}}{(d-a)(d-b)}\right]\nonumber\\
& & \times \left[ \frac{(c-d)~ \eta_a ~e^{-a\tau_1}}{a(a-c)(a-d)}  - \frac{(c-d)~\eta_b ~e^{-b\tau_1}}{b(b-c)(b-d)} + \frac{(a-b) ~\eta_c ~e^{-c\tau_1}}{c(c-a)(c-b)} - \frac{(a-b)~ \eta_d ~e^{-d \tau_1}}{d(d-a)(d-b)}\right]^{N-1}
\end{eqnarray}
In the next section, the above expression for the waiting time distribution for $N$ fluctuating enzyme copies is validated against exact stochastic simulations. The turnover kinetics of fluctuating enzymes is then analysed  in terms of the waiting time distribution and its first and second moments. 

\section{Results and Discussion}

\subsection{Waiting time distribution}

In the presence of enzymatic conformational fluctuations, the analytical expressions for the waiting time distribution for $N$ enzyme copies is given by Eq. (\ref {wtdn-ppm}). This expression is validated against exact stochastic simulations by generating ensembles of typically $10^6$ numerical trajectories of  the CME [Eq. (\ref{cme-ppm})]. Fig. (2) compares the  waiting time distribution obtained from the method of SRP [Eq. (\ref{wtdn-ppm})] against stochastic simulations. An excellent agreement between the analytical expression [Eq. (\ref{wtdn-ppm})] and simulations shows that the method of SRP exactly captures the effects of stochasticity in the turnover kinetics of fluctuating enzymes. 

\begin{figure}[t]
\centering
\includegraphics[scale=.7]{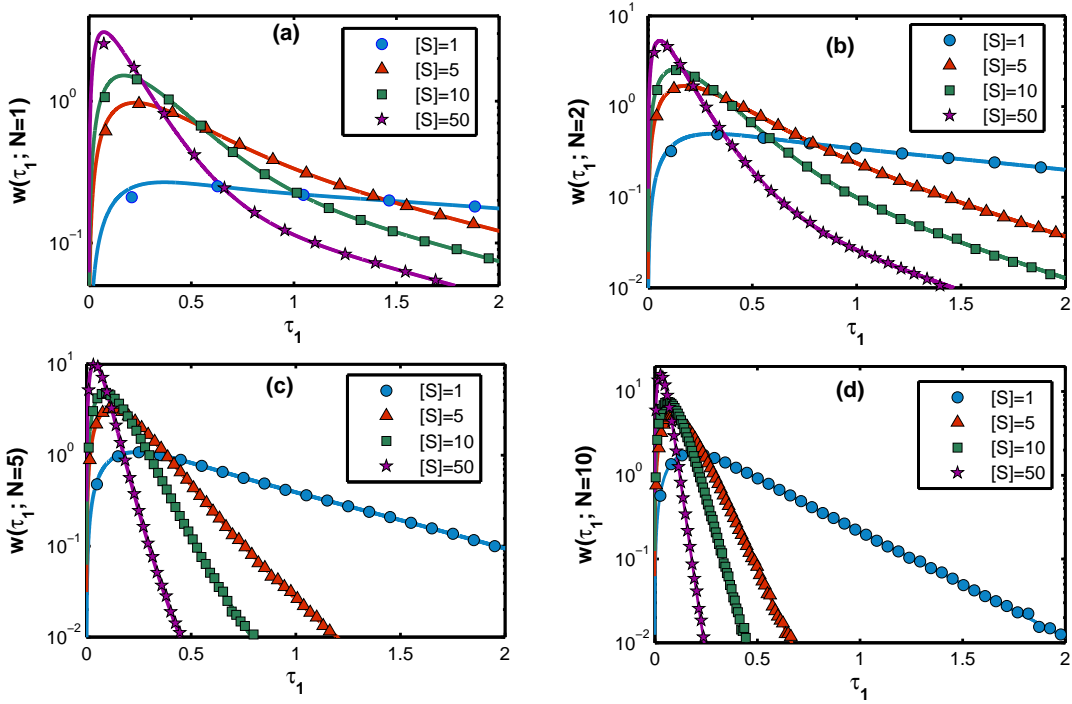}
\vspace{-0.7in}
\caption{Temporal variation of the waiting time distribution (in non-dimensional units) for the parallel-pathway MM mechanism at different $N$. Solid lines are analytical results obtained from Eq. (\ref{wtdn-ppm}) while the symbols are simulation data. For (a) $N=1$ and (b) $N=2$ enzyme(s), the waiting time distribution is mono-exponential at low $[S]$ and multiexponential at high $[S]$. With the increase in the number of enzyme copies, that is, for (c) $N=5$ and (d) $N=50$, the decay at high $[S]$ becomes increasingly monoexponential and the distributions become progressively narrower. Common parameter values in (a)-(d) are $k_{11} = k_{-11} = k_{12} = k_{-12} = 0.5$, $k_{21} = 5$, $k_{22} = 0.1$, $\alpha = \beta = 1, {{\gamma= \delta_{21} = \delta_{22} = 0}}$.}
\end{figure}

\begin{figure}[t]
\centering
\vspace{-0.95in}
\includegraphics[scale=.7]{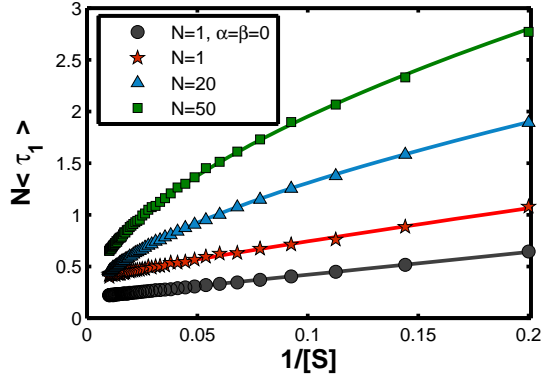}
\vspace{-1.7in}
\caption{Mean waiting time as a function of the reciprocal substrate concentration (in non-dimensional units) plotted in Lineweaver-Burk fashion at different $N$. Solid lines are the first moments of the waiting time distribution, given by Eq. (\ref{wtdn-ppm}), while symbols are simulation data. The parameter values are $k_{11} = k_{-11} = k_{12} = k_{-12} = 0.5$, $k_{21} = 5$, $k_{22} = 0.1$ and $\alpha = \beta = 1, {{\gamma= \delta_{21} = \delta_{22} = 0}}$. The black line represents the classical MM equation and filled circles represent the mean waiting time for a single enzyme in the absence of conformational fluctuations. An excellent agreement between the two shows that the classical MM equation is exactly recovered at the single-molecule level in the absence of conformational fluctuations. Deviations from the classical MM equation occur at all values of $N$ in the presence of conformational fluctuations [red, blue, green symbols and lines].}
\end{figure}

\begin{figure}[t]
\centering
\vspace{-0.45in}
\includegraphics[scale=.7]{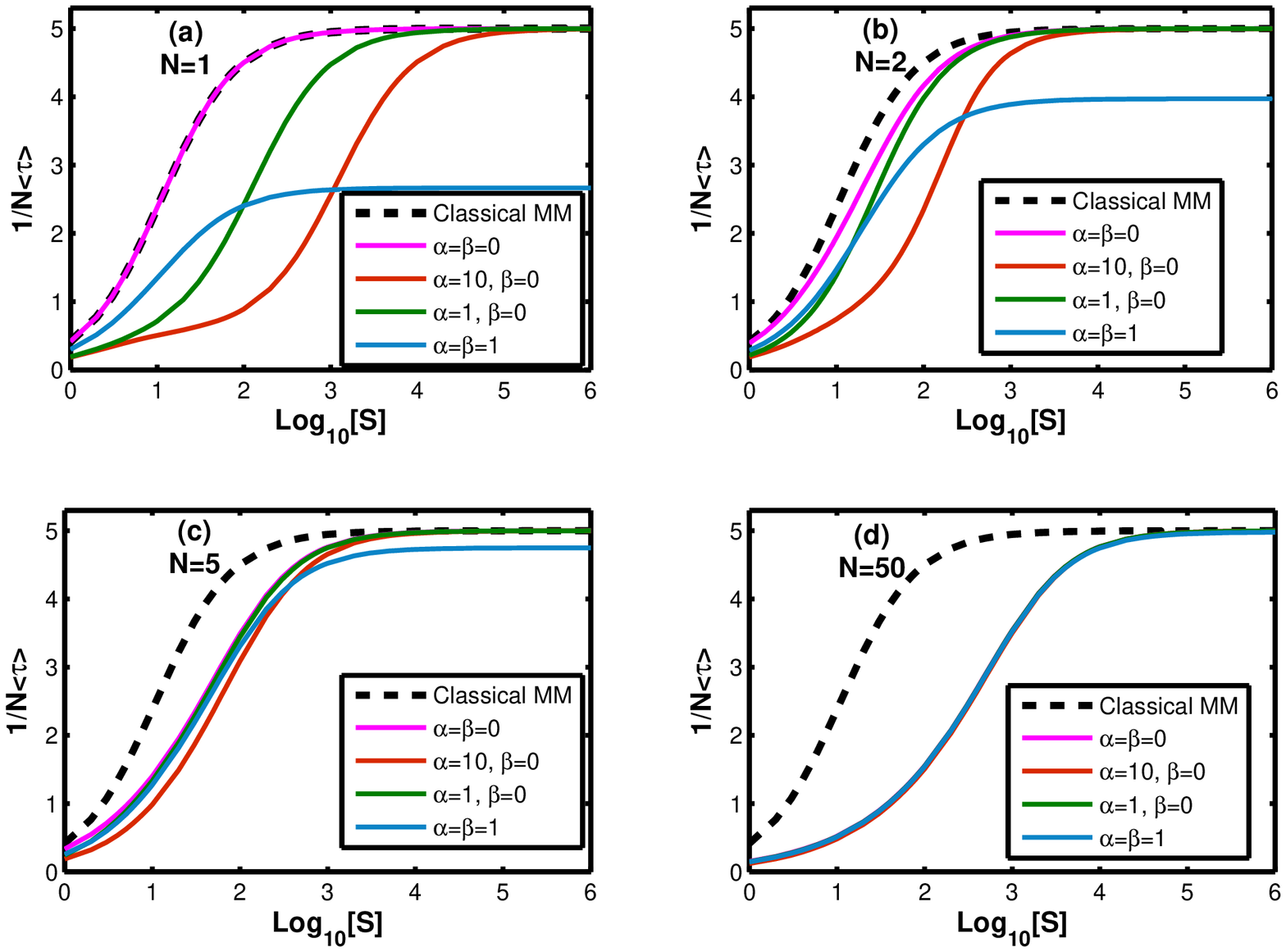}
\caption{Variation of $1/(N \left<\tau_1\right>)$ with substrate concentration (in non-dimensional units) showing the emergence of dynamic cooperativity at mesoscopic number of enzyme copies.  For $N >1$, deviations from the MM equation show a slowing down of the kinetics resulting in negative cooparativity. (a) The effects of conformational fluctuations are clearly discerned at the single enzyme level. For fewer enzymes, (b) and (c) show that the effects of conformational fluctuations are partially suppressed as different curves, representing different time scales of conformational fluctuations, tend to merge with each other. (d)  For large enzyme numbers, different curves collapse onto each other, implying that the effects of conformational fluctuations are completely suppressed. In the latter case, negative cooperativity arises solely due to the discrete nature of reacting species. Common parameter values in (a)-(d) are $k_{11} = k_{-11} = k_{12} = k_{-12} = 0.5$, $k_{21} = 5$, $k_{22} = 0.1, {{\gamma= \delta_{21} = \delta_{22} = 0}}$. }
\end{figure}

Fig. (2) shows the temporal variation of  the waiting time distribution for single and multiple enzyme(s).  For fewer enzymes [Figs.(2a)-(2b)], a broad distribution of waiting times decays monoexponentially at low $[S]$ and multiexponentially at high $[S]$. With the increase in the number of enzymes [Figs. (2b)-(2d)],  the distribution becomes progressively narrower. Also, the decay becomes increasingly steeper and monoexponential. 

At the single enzyme level [Fig. (2a)], the multiexponentiality at high $[S]$ is due to the presence of enzymatic conformational fluctuations, which occur slower on the time scale of the catalytic step, resulting in dynamic disorder in the reaction pathway.  The latter is in agreement with the recent single-enzyme kinetic measurements.$^{12,13}$ To understand the nature of decay for multiple enzyme copies [Figs. (2b)-(2d)], the expression for the waiting time distribution [Eq.(\ref{wtdn-ppm})] can be analyzed for the simple cases of $N=2$ and $N=3$ enzyme numbers.  For $N=2$ enzymes, there are four exponential terms in the second bracket with the effective decay rate constants $a$,  $b$, $c$ and $d$. For $N=3$ enzymes, there are ten exponential terms in the second bracket with the effective decay rate constants that are roughly two times higher than the $N=2$ case. Thus, although the waiting time distribution becomes increasingly multiexponential with the increase in $N$,  the increase in the multiexponential behaviour is coupled to the increase in the magnitude of the effective rate constants for decay. The latter implies steeper decay of the waiting time distribution yielding progressively narrower distributions with $N$. This, combined with a monoexponential decay of waiting time distributions at large $N$, implies that the decay is governed by either a single or multiple rate determining step(s).  These results are further analysed in the next two sections in terms of the first and second moments of the distribution.

\subsection{Mean waiting time and dynamic cooperativity}

The first  moment of the waiting time distribution yields the mean first waiting time, $\left< \tau_1 \right> = \int_0^{\infty} ~ d\tau_1~\tau_1~ w(\tau_1; N)$. For a single enzyme in the absence of enzymatic conformational fluctuations, the first moment of Eq. (\ref{wtd1})  yields $\left< \tau_1 \right> = \frac{k_1[S] + k_{-1} + k_2}{k_1 k_2 [S]}$. The latter shows that $1/\left< \tau_1 \right> = v/[E]_0$ implying that the reciprocal of mean waiting time obtained from the single-enzyme waiting time distribution is related to the (ensemble-average) enzymatic velocity in the classical MM kinetics. Fig. (3) shows the variation of the mean waiting time with the reciprocal of the substrate concentration in the Lineweaver-Burk fashion. Here,  filled markers/symbols represent results obtained from stochastic simulations and lines are analytical results obtained  from the first moment of Eq. (\ref{wtdn-ppm}). For a single enzyme in the absence of enzyme conformational fluctuations [$\alpha = \beta =0$], the mean waiting time obtained from Eq. (\ref{wtd1-ppm}) [filled black circles] is in exact agreement with the classical MM equation [black line]. In the presence of enzyme conformational fluctuations, deviations from the classical MM equation are observed even at the single enzyme level [red line]. The extent of deviation increases with the increase in $N$, but approaches the classical MM equation [black line] in the limit of very high substrate concentration. This limit is analysed in more detail below.

As discussed in the introduction, dynamic cooperativity in monomeric enzymes is described in terms of  a non-Michaelis-Menten kinetic behaviour.$^{1-5}$ Quantified in terms of the variation in $v/[E]_0$ with $[S]$,  a more steep increase in the reaction rate than is allowed by the MM kinetics is termed as the positive cooperativity. A less steep increase in the reaction rate than allowed by the MM equation is referred to as the negative cooperativity. Fig. (4)  captures the effects of dynamic cooperativity in terms of the variation of $1/(N \left<\tau_1\right>)$ with $[S]$  for different enzyme numbers, $N$. 

Fig. (4a) shows the effect of conformational fluctuations on the kinetics of a single enzyme. In the absence of enzymatic conformational fluctuations, the classical MM equation is exactly recovered [pink line]. In the presence of enzymatic conformational fluctuations, deviations from the MM equation are observed in terms of the increase in the mean waiting time with respect to the MM equation. The nature and extent of deviation depend strongly on whether the conformational fluctuations are allowed in both the  enzyme and enzyme-substrate states [blue line] or only in the enzyme state [green and red lines]. In the former case, when the enzymatic ($E$ and $ES$) conformational fluctuations occur slower on the time scale of the catalytic step,  the non-MM behaviour is observed at all $[S]$. In the latter case, the MM equation is recovered at very high substrate concentration. 

The increase in the number of enzymes from single to multiple copies, captured in Figs. (4b)-(4d), shows the emergence of dynamic cooperativity in terms of the non-MM kinetic behaviour. The latter shows strong dependence on the number of enzymes and time scales of enzymatic conformational fluctuations. Interestingly, even a small increase in the number of enzymes from $N=1$ to $N=2$ shows drastic change in the kinetic response associated with dynamic cooperativity. In particular, deviations from the MM equation are observed even when conformational fluctuations are absent. In addition, the MM equation is recovered at relatively lower values of $[S]$ [Fig. (4b)], compared to the $N=1$ case. This shows that pooling of N independent single-enzyme trajectories does not result in N-fold decrease in the mean waiting time as expected from the MM kinetics. Instead, $ \left<\tau_1(N > 1)\right> > \left<\tau_1(N = 1)\right>/N$,  indicating a slowing down of the kinetics due to negative cooperativity.  In the absence of conformational fluctuations, dynamic cooperativity arises solely due to the discrete nature of the reacting species.  With the increase in $N$, the extent of deviation from the MM equation becomes progressively larger [pink line]. This effect due to molecular discreteness stems form the increase in the number of possible reaction pathways with the increase in $N$.  

When conformational fluctuations are included, individual contributions from  enzyme and enzyme-substrate fluctuations towards negative cooperativity can be clearly discerned for fewer enzymes [blue, red and green lines in Fig. (4b)]. The increase in the number of enzymes suppresses the effects  of conformational fluctuations. This is captured in Fig. (4c), in terms of the variation  of $1/(N \left<\tau_1\right>)$ with $[S]$, where the kinetic behaviour in the presence of conformational fluctuations represented by blue, red, green lines partially merges with the pink line representing the absence of conformational fluctuations. At large $N$,  the collapse becomes complete and all lines merge with each other [Fig. (4d)]. These results indicate that dynamic cooperativity at small $N$ is the result of the discrete nature of fluctuating enzymes. At large $N$, when the conformational fluctuations are suppressed, dynamic cooperativity stems solely from the molecular discreteness.

At very high substrate concentration, the MM kinetics is exactly recovered for large enzyme numbers [Figs. (4d)]. For fewer enzymes, however, convergence to the MM equation depends on the time scales of  enzymatic conformational fluctuations. When enzymatic conformational fluctuations occur slower on the time scale of the catalytic step, the MM kinetics is not recovered even at very high $[S]$ [blue curve in Fig. (4b)-(4c)]. When conformational fluctuations are allowed only in the enzyme state, the MM equation is exactly recovered at high $[S]$. The latter occurs irrespective of the time scale of conformational fluctuations, which can be faster or slower than the catalytic step, depicted by the red and green lines in Fig. (4b)-(4c) respectively.  These results will be further analysed in the next subsection in terms of the dimensionless variance of the waiting time distribution, termed as the randomness parameter.

\subsection{Randomness parameter - a quantitative measure of temporal fluctuations}

The randomness parameter is the dimensionless ratio of the variance to the squared-mean, $r =  \frac{\left< \tau_1^2 \right> - \left< \tau_1 \right>^2}{\left< \tau_1 \right>^2}$. It quantifies the magnitude of temporal fluctuations in the underlying reaction pathway. Fig. (5) shows the variation of the randomness parameter with  $N$ for a given value of $[S]$. The presence and absence of enzyme conformational fluctuations correspond to the stochastic kinetics based on the parallel-pathway [Eq. (\ref{ppm})] and single-pathway [Eq. (\ref{mmm})]  MM mechanisms, respectively. 

In the presence of conformational fluctuations, the randomness parameter is greater than one for fewer enzymes. This reflects a broad distribution of waiting times with multiexponential decay, depicted in Figs. (2a)-(2b). For conformational fluctuations occurring slower on the time scale of the catalytic step,  the value of $r > 1$ signifies multiple competing steps in the parallel-pathway mechanism $^{12-14,18}$. This effect due to dynamic disorder results in substantial temporal fluctuations for small number of enzymes. With the increase in the enzyme numbers, the value of the randomness parameter switches to $r \leq 1$. The latter indicates a narrow distribution of waiting times which decay monoexponentially [Figs. (2c)-(2d)]. In addition, the value of $r = 1$ or $r < 1$ indicate a sequential (single-pathway) MM mechanism with single or multiple rate determining step(s), respectively.$^{18,51-53}$

Fig. (5), thus, shows that the randomness parameter for the parallel- and single-pathway MM mechanisms completely merge with each other at large $N$, both of them yielding an identical value of $r < 1$.  This, on the one hand, implies that monoexponential decay of the waiting time distributions in Fig. (2d) is associated with the multiple rate determining steps in the (single-pathway) MM mechanism. On the other hand, it quantitatively shows that conformational fluctuations are completely suppressed at large $N$. Together, these results imply  that the turnover kinetics of fluctuating enzyme based on the parallel-pathway MM mechanism  switches over to the single-pathway MM mechanism with the increase in the number of enzymes. In the former case the stochasticity is due to the combined effect of enzymatic conformational fluctuations and molecular discreteness. In the latter case, stochasticity stems solely from the molecular discreteness.

\begin{figure}[t]
\centering
\hspace*{-3.8em}
\vspace{-1.35in}
\includegraphics[scale=.8]{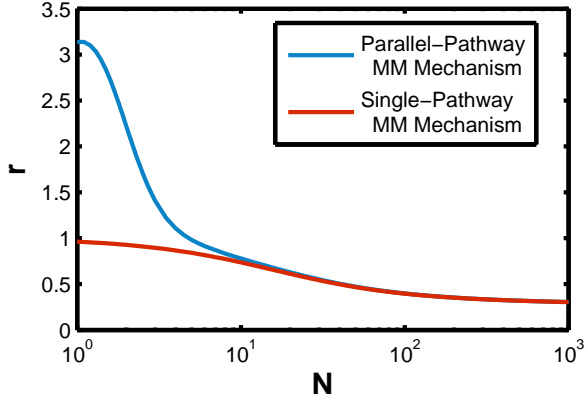}
\vspace{-.35in}
\caption{Variation of the randomness parameter with $N$ for parallel- and single pathway MM mechanisms. When conformational fluctuations occur slower on the time scale of the catalytic step, $r > 1$ for fewer enzymes and $r < 1$ for more enzymes.  The randomness parameter for the parallel- and single-pathway MM mechanisms completely merge with each other at large $N$ implying that the parallel-pathway MM mechanism switches over to the single-pathway mechanism at large $N$. The parameter values are $k_{11} = k_{-11} = k_{12} = k_{-12} = 0.5$, $k_{21} = 5$, $k_{22} = 0.1$, $\alpha = \beta = 1, {{\gamma= \delta_{21} = \delta_{22} = 0}}$ and $[S]=500$.}
\end{figure}

\begin{figure}[t]
\centering
\hspace*{-2.8em}
\vspace{-0.95in}
\includegraphics[scale=.85]{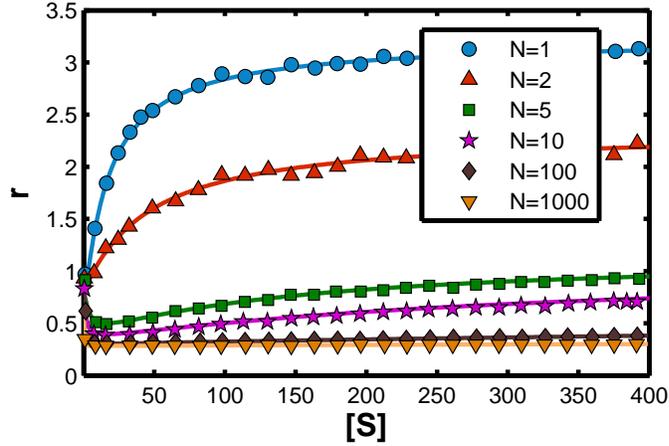}
\vspace{-.75in}
\caption{Variation of the randomness parameter with $[S]$ (in non-dimensional units) for conformational fluctuations occurring slower on the time scale of the catalytic step. For single and fewer enzyme(s), $r > 1$ due to dynamic disorder in the parallel-pathway mechanism. Since conformational fluctuations are allowed in both the enzyme and enzyme-substrate state, the value of $r  > 1$  at high substrate concentration.  For more enzymes, the effects of enzyme conformational fluctuations is suppressed, and $r < 1$ for all values of $[S]$. Solid lines represent the dimensionless variance of the waiting time distribution, given by Eq. (\ref{wtdn-ppm}), while symbols are simulation data. The parameter values are $k_{11} = k_{-11} = k_{12} = k_{-12} = 0.5$, $k_{21} = 5$, $k_{22} = 0.1$ and $\alpha = \beta = 1, {{\gamma= \delta_{21} = \delta_{22} = 0}}$.}
\end{figure}

{Interestingly, a recent work finds the randomness parameter to be greater than unity at all substrate concentrations.$^{54}$ Since the latter uses a method different from the CME approach, it is not possible for us to make a direct comparison between the results obtained from two different theoretical approaches.   It is possible, however, to  make a direct comparison  between the randomness parameter obtained using the method of SRP [Eq. (\ref{wtdn-ppm})] and stochastic simulations that solves the CME [Eq. (\ref{cme-ppm})] exactly.  The result of this comparison is shown in Fig (6), which provides an excellent agreement between the randomness parameter obtained using the SRP method [lines] and simulation data [symbols] for all values of substrate concentration.} For fewer enzymes, $r >1$ for all values of $[S]$ due to dynamic disorder. The value of $r >1$  at high substrate concentration stems from the presence of conformational fluctuations in both the enzyme and enzyme-substrate state, which occur slower on the time scale of the catalytic step. For more enzymes, the effects of enzyme conformational fluctuations is suppressed, and $r < 1$ for all values of $[S]$. The crossover from $r>1$ to $r<1$, once again, suggests that  the parallel-pathway MM mechanism switches  over to the single-pathway MM mechanism with the increase in $N$. While $r >1$ suggests substantial conformational fluctuations at low enzyme numbers, the value of $r < 1$  confirms  that conformational fluctuations are significantly suppressed at large enzyme numbers. 

Fig. (6), thus, shows that deviations from the MM equation for small number of enzymes at very high $[S]$, captured  by the blue line in Figs. (4a)-(4c), is the result of significant conformational fluctuations  amounting to $r > 1$ at all $[S]$. The increase in $N$ suppresses these fluctuations resulting in $r < 1$ at all $[S]$. The latter implies that the MM equation is exactly recovered at very high $[S]$, as captured by the blue line in Fig. (4d).


\section{Summary and Conclusion}

In this work, we have presented a bottom up approach based on the superposition of renewal processes (SRP). This approach exploits the renewal nature of the waiting time distribution of a single enzyme to obtain the waiting time distribution of $N$ enzyme copies as a single pooled output. In the absence of conformational fluctuations, the waiting time distribution obtained from the method of SRP yields the same {\it exact} result as the generating function method. In the presence of conformational fluctuations, when the analytical solution using the generating function method is difficult, the waiting time distribution obtained from the method of SRP is in exact agreement with stochastic simulations for all values of $N$ and $[S]$. 

The  waiting time distribution shows multiexponential decay for fewer enzymes. The multiexponentiality arises due to the combined effect of conformational fluctuations and molecular discreteness.  The increase in the number of enzymes, however, suppresses the effect of conformational fluctuations such that the decay becomes progressively monoexponential. This behaviour is further quantified by the randomness parameter, $r$, which shows a crossover from a value of greater than one to less than one with the increase in the number of enzymes. The value of $r > 1$  arises due to multiple competing steps in the parallel-pathway mechanism. This effect due to dynamic disorder results in substantial temporal fluctuations at smaller $N$.  At larger $N$, the randomness parameter for the parallel-pathway MM mechanism is identical with the  single-pathway MM mechanism, implying that the underlying turnover kinetics switches from the parallel-pathway to single-pathway with the increase in $N$. The temporal fluctuations in the latter stems solely from the discrete nature of reacting species. 

The reciprocal of the mean waiting time exactly recovers the MM equation for a single enzyme with no conformational fluctuations.  When conformational fluctuations are allowed,  dynamic cooperativity in multiple enzyme copies emerges in terms of a non-Michaelis-Menten kinetic behaviour.  The latter shows a slowing down of the MM kinetics resulting in negative cooperativity. For fewer enzymes, dynamic cooperativity emerges due to the combined influence of enzymatic conformational fluctuations and molecular discreteness. In particular, the distinct contributions of enzyme and enzyme-substrate fluctuations can be clearly discerned  in terms of the extent of deviation from the MM kinetics. The increase in the number of enzymes suppresses the effect of enzymatic conformational fluctuations such that dynamic cooperativity emerges solely due to the discrete changes  in the number of reacting species. In the limit of very large substrate concentration, the MM equation is exactly recovered for large enzyme numbers as the stochastic turnover kinetics approaches the deterministic behaviour. For small enzyme numbers, with conformational fluctuations in both the enzyme and enzyme-substrate state that occur slower on the time scale of the catalytic step, deviations from the MM equation are observed even at very large substrate concentration.

The detailed analysis of the waiting time distribution, the mean waiting time and the randomness parameter, together, confirm that the turnover kinetics of fluctuating enzyme based on the parallel-pathway MM mechanism  switches over to the single-pathway MM mechanism with the increase in the number of enzymes. A close connection between dynamic cooperativity and dynamic disorder for small enzyme numbers seem to emerge naturally from the stochastic kinetic approach presented here.

As mentioned in the introduction, enzyme cooperativity is traditionally linked to enzymes with multiple binding sites, where binding of a substrate at a site can influence the time scale of substrate binding at another site. The latter is the result of interaction between sites. The present formalism, on the other hand, considers $N$ independent and identical copies of a monomeric enzyme to show the emergence of dynamic cooperativity in terms of deviation from the MM equation. The generality of the present formalism allows us to make a correspondence between the kinetics of $N$ independent and identical copies of a monomeric enzyme with the kinetics of a single enzyme with $N$ independent and identical sites. Depending on the number of sites, therefore, cooperativity in a single enzyme with identical and independent multiple sites can be shown to emerge purely from the stochastic effects - enzymatic conformational fluctuations and molecular discreteness. This shows that, stochastic effects alone, are sufficient to produce cooperativity and interactions between sites are not necessary.  This, then, presents an opportunity to study problems in single-molecule reactions catalyzed by an enzyme with multiple binding sites$^{12,13}$ in order to understand, separately, the roles of stochasticity and interaction. 
\eject

\section{supplementary Information}
See Supplementary Information for details of Simulation and analytical methods used.

\begin{acknowledgments}
AK acknowledges the financial support from the Council of Scientific and Industrial Research (CSIR), Government of India.
\end{acknowledgments}


\end{document}